\documentclass{article}

    \PassOptionsToPackage{numbers, compress}{natbib}
\usepackage[preprint]{neurips_2024}

\usepackage[utf8]{inputenc} % allow utf-8 input
\usepackage[T1]{fontenc}    % use 8-bit T1 fonts
\usepackage{hyperref}       % hyperlinks
\usepackage{url}            % simple URL typesetting
\usepackage{booktabs}       % professional-quality tables
\usepackage{amsfonts}       % blackboard math symbols
\usepackage{nicefrac}       % compact symbols for 1/2, etc.
\usepackage{microtype}      % microtypography
\usepackage{xcolor}         % colors
\usepackage{graphicx}
\usepackage{amsmath}
\usepackage{color,soul}
\title{Inverse receptive field attention for naturalistic image reconstruction from the brain}

\author{
    \textbf{Lynn Le}\textsuperscript{1}, 
    \textbf{Thirza Dado}\textsuperscript{1},
    \textbf{Katja Seeliger}\textsuperscript{1},
    \textbf{Paolo Papale}\textsuperscript{2}, 
    \textbf{Antonio Lozano}\textsuperscript{2}, \\
    \textbf{Pieter Roelfsema }\textsuperscript{2,3,4,5}, 
    \textbf{Yağmur G{\"u}{\c{c}}l{\"u}t{\"u}rk}\textsuperscript{1}, 
    \textbf{Marcel van Gerven }\textsuperscript{1},
    \textbf{Umut G{\"u}{\c{c}}l{\"u}}\textsuperscript{1}\\
    \vspace{1em} % Space between authors and affiliations
    \begin{tabular}{@{}c@{}}
        \small \textsuperscript{1} Donders Institute for Brain, Cognition and Behaviour, Radboud University, Nijmegen, Netherlands\\
        \small \textsuperscript{2} Netherlands Institute for Neuroscience, Amsterdam, Netherlands \textsuperscript{2}  Department of Integrative Neurophysiology,\\
        \small  Centre for Neurogenomics and Cognitive Research, Vrije Universiteit \textsuperscript{5} Laboratory of Visual\\
        \small Brain Therapy, Paris, France \textsuperscript{4} Department of Psychiatry, Amsterdam UMC, University of Amsterdam \\ \\
        \small \dag Correspondence Email: lynn.le@donders.ru.nl
    \end{tabular}
} 

\begin{document}

\maketitle

\begin{abstract}

Visual perception in the brain largely depends on the organization of neuronal receptive fields. Although extensive research has delineated the coding principles of receptive fields, most studies have been constrained by their foundational assumptions. Moreover, while machine learning has successfully been used to reconstruct images from brain data, this approach faces significant challenges, including inherent feature biases in the model and the complexities of brain structure and function. In this study, we introduce an inverse receptive field attention (IRFA) model, designed to reconstruct naturalistic images from neurophysiological data in an end-to-end fashion. This approach aims to elucidate the tuning properties and representational transformations within the visual cortex. The IRFA model incorporates an attention mechanism that determines the inverse receptive field for each pixel, weighting neuronal responses across the visual field and feature spaces. This method allows for an examination of the dynamics of neuronal representations across stimuli in both spatial and feature dimensions. Our results show highly accurate reconstructions of naturalistic data, independent of pre-trained models. Notably, IRF models trained on macaque V1, V4, and IT regions yield remarkably consistent spatial receptive fields across different stimuli, while the features to which neuronal representations are selective exhibit significant variation. Additionally, we propose a data-driven method to explore representational clustering within various visual areas, further providing testable hypotheses.

\end{abstract}

\section{Introduction}
Natural scenes are complex and cluttered, often containing both crucial and irrelevant stimuli. Perception of these scenes largely depends on \textit{where} attention is directed, impacting \textit{what} is perceived. Attention in visual scenes, whether spatial or feature-based, helps prioritize neural resources, enabling focus on specific aspects of the environment \cite{posner1988structures, posner1990attention, desimone1995neural, serences2006selective, galashan2017differences}. Due to the visual system's limited capacity, neural activity is biased toward attended locations, enhancing the representation of more salient objects \cite{ungerleider2000mechanisms}. Research has shown that higher ventral stream regions encode increasingly invariant representations of objects, filtering out surrounding clutter \cite{braun2001visual, itti2001computational}. 

The integration of attention mechanisms into neural coding models plays an important role by creating an information bottleneck, which allows only the most salient objects to be represented in deeper layers \cite{khosla2020neural}. This selective representation significantly enhances the accuracy of predicting neural responses to naturalistic stimuli \cite{khosla2020neural}, providing a strong foundation for extending these models to more complex tasks, such as the reconstruction of real-world images from brain activity. Reconstructing complex real-world images directly from brain data offers deeper insights into neural computation and the brain's interpretation of real-world scenarios \citep{sonkusare2019naturalistic, Shen2019end, nishimoto2011reconstructing, le2022brain2pix}. However, current methods often rely on nonlinear representations of data derived from large, pretrained generative models \citep{seeliger2018generative, dado2024brain2gan, takagi2023high, scotti2024reconstructing, fang2020reconstructing, chen2024cinematic, lin2022mind, shen2019deep}. While these approaches are promising due to their simplicity, they can introduce biases related to their objective function and datasets, limiting their ability to capture the intrinsic complexities of the brain’s encoding mechanisms. To address these challenges, our study introduces a novel end-to-end convolutional neural network (CNN) that incorporates spatial and feature-based selective attention, trained solely on neural data. 

We introduce a novel inverse receptive field attention (IRFA) model designed to visualize the dynamics of neurons when switching between spatial and feature-based attention. In an idealized sense, an inverse receptive field (IRF) for each pixel would correspond to a fixed subset of neurons that specifically respond to that location, creating a direct pixel-to-neuron mapping. However, in practice, this mapping is implemented as a soft assignment to capture the continuous nature of neural interactions across the visual field.

This model generates selective attention maps and co-trains with a fully convolutional reconstruction model trained on V1, V4, and IT of macaque data. Specifically, the IRFA model incorporates an attention component that disentangles spatial and feature selectivity of neuronal representations. This component assigns an inverse receptive field (IRF) map to each pixel before entering a U-NET model, thereby elucidating the visual selective attention across both location and feature dimensions. The model operates end-to-end and is trained on complex natural images to accurately identify spatial and feature-based receptive fields under naturalistic conditions.

The introduction of the IRFA model contributes several advancements to the field. First, it demonstrates that incorporating feature-based selective attention, alongside spatial attention, into brain representations significantly enhances the quality of natural image reconstruction from evoked brain responses in the V1, V4, and IT regions of the macaque.

Second, our model introduces a novel inverse pixel-wise attention mechanism, where multihead embeddings, positional encodings, and a transpose attention mechanism are integrated to estimate inverse receptive fields dynamically. Feature inverse receptive fields (IRFs) refer to the attention mechanisms over specific grid channels or "features" that represent encoded aspects of the visual stimulus. Each feature corresponds to a channel in the virtual grid that captures unique details of the encoded information. This approach transforms the neuron-to-pixel mapping problem into an image-to-image translation task, allowing predicted output pixels to direct attention back to input electrodes through learned inverse queries. This adaptation of self-attention, specifically for neural decoding, establishes a robust pixel-to-neuron mapping, offering a refined approach to image reconstruction.

Third, our model provides evidence that a selective attention network can be co-trained by direct supervision from image reconstruction tasks, operating entirely end-to-end. By combining inverse receptive field mapping with Brain2Pix, our approach significantly extends this framework for enhanced accuracy and interpretability in neural data analysis.

Fourth, it reveals that while spatial inverse receptive fields (spatial IRFs) within our model remain consistent across different stimuli, feature receptive fields (feature IRFs) vary, indicating a dynamic response to varying neural stimuli. This observation provides novel insights into neuronal tuning properties that have not been previously documented.

Finally, IRFA suggests that in-silico experiments on neural reconstruction models can generate new hypotheses and offer a data-driven framework for developing broad, generalizable theories about neural information processing. For instance, a DBSCAN analysis of t-SNE output shows that spatial maps form 15 clusters corresponding to electrode arrays, while feature-based fields reveal potential attention to 83 distinct features across 100 test stimuli.

Taken together, IRFA provides a new approach to elucidate mechanisms of brain function from neural response data, advancing both the technical and theoretical understanding of neural decoding.

% \begin{itemize}
%     \item It demonstrates that incorporating feature-based selective attention, in addition to spatial attention, into brain representations can significantly enhance the quality of natural image reconstruction from evoked brain responses in the V1, V4, and IT regions of the macaque.
%     \item It provides evidence that a selective attention network can be co-trained by direct supervision from image reconstruction tasks, operating entirely end-to-end.
%     \item It reveals that while the spatial inverse receptive fields (spatial IRFs) within our model remain consistent across different stimuli, feature receptive fields (feature IRFs) vary, indicating a dynamic response to varying neural stimuli. This observation provides novel insights into neuronal tuning properties that have not been previously documented.
%     \item It proposes that in-silico experiments on neural reconstruction models can generate new hypotheses and offer a data-driven framework for developing broad, generalizable theories about neural information processing. For instance, a DBSCAN analysis of t-SNE output shows that spatial maps form 15 clusters corresponding to electrode arrays, while feature-based fields reveal potential attention to 83 distinct features across 100 test stimuli.
% \end{itemize}

\section{Preliminaries}
\subsection{Naturalistic image reconstruction}
The goal of neural decoding is to infer perceptual or behavioral information from neural activity. We focus on reconstructing visual stimuli from multi-unit activity (MUA), recorded with micro-electrode arrays placed in the V1, V4, and IT regions of the macaque. We define a visual stimulus as a tensor \( S \in \mathbb{R}^{c \times h \times w} \), with \( c \) representing color channels and \( h \times w \) denoting the height and width of the pixel grid. Given neural recordings \( R \in \mathbb{R}^{m} \), where \( m \) is the number of electrode arrays, the challenge is to recover \( S \) through a decoding function \( d \):
\[
S = d(R) \,.
\]

This function \( d \) encompasses complex dynamics given the nonlinear and high-dimensional nature of the brain's encoding of visual information. As mentioned before, traditional neural decoding approaches often fall short in capturing the intrinsic complexities of the brain's encoding mechanisms. Thus, we employ our IRF reconstruction model that leverages the concept of receptive fields and their inverses, combined with advanced deep learning architectures to approximate \( d \).

\subsection{Receptive Fields}
In the context of visual neurons, the receptive field refers to the specific region of the visual field that elicits a response from the neuron~\cite{Hubel1959}. Mathematically, for a given neuron \( i \), its receptive field can be defined as a subset of pixels \( \mathcal{R}_i \) in the stimulus space. Each neuron, therefore, selectively responds to stimuli presented within its receptive field, effectively acting as a filter that spatially localizes the neuron's sensitivity.

Let \( e \) denote the encoding function that maps a visual stimulus \( S \) to neural activity across time and microelectrodes. The recorded activity of neuron \( i \), represented by \( a_{i} \), can be predominantly influenced by the portion of the stimulus within its receptive field. That is, 
\[
a_{i} \approx e(S_{|\mathcal{R}_i}) 
\]
where \( S_{|\mathcal{R}_i} \) denotes the restriction of the visual stimulus to the receptive field of neuron \( i \). This approximation accounts for the fact that neurons aggregate and transform visual information within their receptive fields, without capturing the full complexity of neural encoding mechanisms.

\subsection{Inverse receptive fields}
The inverse of the concept of receptive fields leads us to the novel idea of the inverse receptive field (IRF). For a given pixel in the visual field, defined by its coordinates \( (x, y) \), the inverse receptive field is the set of neurons whose receptive fields include that pixel. Formally, for pixel \( p \), the IRF is given by:
\[
\mathcal{IRF}_p = \{i \mid p \in \mathcal{R}_i\} \,.
\]
This suggests that each pixel in the visual field can be associated with a unique set of neurons that are attentive to visual information present at that pixel's location. The concept of the inverse receptive field captures the collective sensitivity of the visual neuron population to each point in visual space, providing a reversed perspective on the spatial encoding of visual information in the brain.

By leveraging the inverse receptive field, we are able to invert the encoding process, focusing instead on how the distributed neural activity conveys information about specific visual field locations. This lays the foundation for reconstructing the visual stimulus by aggregating the contributions from neurons that share sensitivity to particular pixels, employing a spatial transformation that effectively maps neural representations back to visual space.

In essence, the reconstruction of a pixel's value in the visual stimulus leverages the fact that the relevant information is encoded within the collective activity of neurons defined by the pixel's IRF, enabling a spatially resolved reversal of the brain's encoding of visual stimuli. This insight into the spatially distributed encoding process facilitates a structured approach to neural decoding, specifically through the implementation of inverse receptive field mapping.

\begin{figure}[!ht]
  \centering
  \includegraphics[width=\textwidth]{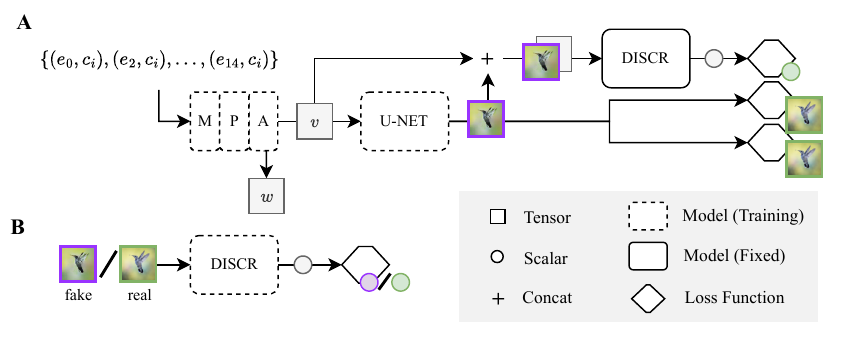}
\caption{Schematic of our IRF reconstruction model comprising two forward passes and training five components simultaneously. \textbf{A}: Input from 15 electrode arrays—7 from V1, 4 from V4, and 4 from IT, with \(n\) channels each—is processed through multi-head embedding (M), positional encoding (P), and transpose attention (A) to create an attention map \(w\) and a weighted map \(v\). The map \(v\) is used by the U-NET for naturalistic image reconstruction, evaluated using discriminator, VGG, and L1 losses. The map \(w\) will be analyzed for visualizing the spatial IRFs and feature IRFs. \textbf{B}: The second forward pass involves discriminator training on U-NET `fake' outputs versus `real' stimuli.}
  \label{fig:architecture}
\end{figure}
\section{Methods}
\subsection{Inverse receptive field attention model}
Our inverse attention mechanism maps brain responses onto a virtual spatial grid with multiple channels, each channel representing distinct features of the neural representation. In this context, "features" refer to the individual channels in the embedding space, where each channel captures specific aspects of the encoded information at each grid point. Each point on this grid corresponds to a spatial location in the visual field, while each feature (channel) encodes details relevant to that location. This configuration allows the model to generate both spatial receptive fields (over grid locations) and feature receptive fields (over channels) for each electrode. By selectively attending to the input electrodes, each grid point learns spatial and feature-specific attention patterns, which are then used to reconstruct the visual stimulus. 

\paragraph{Multi-head embedding}

The multi-head embedding transforms neural recordings from each microelectrode array into a set of feature embeddings. Given input data \( X = (x_1, x_2, \ldots, x_N) \), where each \( x_i \in \mathbb{R}^{b \times n_i} \) represents recordings from an electrode array, we treat each array as a distinct input head. Each head is processed with an affine transformation:
\[
E_h = \text{Affine}_h(R_h)
\]
to create embeddings \( E_h \in \mathbb{R}^{\frac{C}{2}} \), where \( C \) is a critical hyperparameter defining the feature space dimensionality. These embeddings are combined into a tensor \( E \in \mathbb{R}^{N \times H \times \frac{C}{2} \times 1 \times 1} \), capturing relevant spatial and feature information for further processing. 

\paragraph{Positional encoding}

To incorporate spatial information into the embeddings, we introduce positional biases in both \( x \) and \( y \) dimensions. These positional encodings, \( P_x \) and \( P_y \), are learnable parameters defined as:
\[
P = \text{Cat}\left( \left[E + P_x, E + P_y \right], \text{dim}=2 \right)
\]
where \( P_x \in \mathbb{R}^{1 \times 1 \times \frac{C}{2} \times 1 \times I} \) and \( P_y \in \mathbb{R}^{1 \times 1 \times \frac{C}{2} \times I \times 1} \). This concatenation results in the positional encoding tensor \( P \in \mathbb{R}^{N \times H \times C \times I \times I} \), effectively embedding spatial information within each feature representation.

\paragraph{Transpose attention}

The transpose attention component utilizes the positional encoding \( P \) to compute attention scores and aggregate information across the visual field. Key and value vectors are generated as:
\[
K = \text{Conv2d}(P, k=1), \quad V = \text{Conv2d}(P, k=1)
\]
where \( \text{Conv2d} \) represents a 2D convolution. The attention scores \( W \) are computed as:
\[
W = \text{Softmax}\left( \frac{ \text{Conv2d}(K) }{ \sqrt{C \times k \times k} } \right)
\]
where \( k \) is the kernel size, and the scaling factor \( \sqrt{C \times k \times k} \) stabilizes values in high-dimensional space. The attention output \( A \) is computed by weighting the value vectors:
\[
A = \sum_{h=1}^{H} W_h \odot V_h
\]
where \( \odot \) denotes element-wise multiplication. The resulting tensor \( A \in \mathbb{R}^{N \times C \times I \times I} \) provides a spatially aware, feature-rich representation of the neural activity, mapping it onto the visual field and setting the stage for the U-NET-based reconstruction.

\paragraph{U-NET model}
The final weighted attention map (\(v\)) is used to train the U-NET model, as shown in Figure~\ref {fig:architecture}A. The map \(v\) is passed into the U-NET layers, which consists of 2DConvolutional layers Transposed2DConvolutional layers, and skip layers. Skip layers consist of 2DConvolutional, BatchNorm2D, relu, its previous layer, deconv, and batchnorm. The configuration of each layer is reported in Table~\ref{table:unet_flow}. 

\begin{table}[!ht]
\centering
\caption{Layers of the U-NET model.}
\label{table:unet_flow}
\resizebox{!}{57pt}{%
\begin{tabular}{lll}
\hline
\textbf{Layer} & \textbf{Shape} & \textbf{Configurations} \\ \hline
Input & $96 \times 96 \times 15$ & - \\
Conv2d 1 & $48 \times 48 \times 64$ & kernel\_size=4, stride=2, padding=1 \\
LeakyReLU 1 & $48 \times 48 \times 64$ & negative\_slope=0.2, inplace=True \\
Identity & $48 \times 48 \times 512$ & Identity parameters: count=512, depth=512 \\
Skip 1 (Conv2d $\rightarrow$ ConvTranspose2d) & $48 \times 48 \times 512$ & See detailed breakdown \\
Skip 2 (Conv2d $\rightarrow$ ConvTranspose2d) & $48 \times 48 \times 256$ & See detailed breakdown \\
Skip 3 (Conv2d $\rightarrow$ ConvTranspose2d) & $48 \times 48 \times 128$ & See detailed breakdown \\
Skip 4 (Conv2d $\rightarrow$ ConvTranspose2d) & $48 \times 48 \times 64$ & See detailed breakdown \\
ConvTranspose2d & $96 \times 96 \times 3$ & kernel\_size=4, stride=2, padding=1 \\
Sigmoid & $96 \times 96 \times 3$ & - \\ 
Output & $96 \times 96 \times 3$ & - \\ 

\hline
\end{tabular}%
}
\end{table}

The U-NET's output is compared with the target using three losses (an adversarial loss, a feature loss (VGG) and an L1 loss). The adversarial loss is a discriminator that is trained in parallel of the reconstruction model, which consists of 5 convolutional layers (see Fig.~\ref{fig:architecture}B). The feature loss uses the full set of convolutional layers of the VGG model. The model was implemented in Pytorch and optimized with Adam in 400 epochs with a batch size of 8. The implementation of the model can be found in the source code \footnote{\url{https://github.com/neuralcodinglab/IRFA}}.

\subsection{Experiments} \label{sec:experiments}
We systematically trained the model to reconstruct images using sets of 4, 16, 32, and 64 learnable attention maps. This approach allows us to evaluate the optimal number of features required for effective reconstruction. The quality of the reconstructions is quantitatively compared with a baseline model, based on the end-to-end reconstruction model of Shen et al.~\cite{Shen2019end}. Additionally, we assess the consistency of the spatial and feature inverse receptive fields (feature IRFs) across different stimuli by computing the standard deviation. Lastly, we employ a data-driven approach to visualize changes in feature RFs, using t-SNE for dimensionality reduction to explore whether electrodes exhibit preferences for certain features.

\paragraph{Experimental data}
The THINGS database contains a rich collection of naturalistic object images, featuring 26,000 images across 1,854 diverse categories \footnote{\url{https://things-initiative.org}}. A total of 25,248 images from this database were shown to a macaque monkey, using 12 images from each category, as detailed in Hebart et al., 2019 \cite{hebart2019things}.

The experimental setup involved a macaque monkey equipped with a 960-channel neural implant, composed of 15 functioning Utah arrays~\cite{chen2020shape, chen20221024}. %However, electrode array \#6 is broken, so for the rest of the paper, consider 15 electrodes.  
These arrays were distributed as follows: seven in V1, four in V4, and four in IT. The images were divided into training and test sets, presented randomly. Each category in the training set was shown once, while the test set included 100 images, each displayed 30 times.
More explicitly, the number of channels in each of the arrays are as follows:
\begin{align*}
    \text{V1 Arrays:} & \quad x_1 \in \mathbb{R}^{b \times 43}, \; x_2 \in \mathbb{R}^{b \times 59}, \; x_3 \in \mathbb{R}^{b \times 37}, \; x_4 \in \mathbb{R}^{b \times 44}, \\
    & \quad x_5 \in \mathbb{R}^{b \times 50}, \; x_6 \in \mathbb{R}^{b \times 2}, \; x_7 \in \mathbb{R}^{b \times 48}, \\
    \text{V4 Arrays:} & \quad x_8 \in \mathbb{R}^{b \times 43}, \; x_{9} \in \mathbb{R}^{b \times 44}, \; x_{10} \in \mathbb{R}^{b \times 53}, \; x_{11} \in \mathbb{R}^{b \times 29}, \\
    \text{IT Arrays:} & \quad x_{12} \in \mathbb{R}^{b \times 17}, \; x_{13} \in \mathbb{R}^{b \times 24}, \; x_{14} \in \mathbb{R}^{b \times 59}, \; x_{15} \in \mathbb{R}^{b \times 24}.
\end{align*}

% Note: Array 6 is broken and excluded from \( X \).

During the experiments, the monkey was trained to fixate on a red dot against a gray background for 300ms. This was followed by a rapid presentation of four images, each lasting 200ms with an inter-trial interval of 200ms. The images, measuring 500 by 500 pixels, were shifted 100 pixels toward the lower right of the fovea. If the monkey maintained fixation throughout the sequence, it received a juice reward.

Multi-unit activity (MUA) responses were recorded, representing the collective spiking activity of local neuron networks \cite{burns2010comparisons}. The data preprocessing included averaging the raw signals to reduce noise and normalizing by subtracting the mean response of all test trials for the day from each trial and channel. This was followed by division by the standard deviation of these trials. To assess data reliability, Pearson correlation coefficients were calculated for all pairs of the 100 test images, yielding 435 coefficients per electrode channel \((30 \times (30-1) / 2)\). These correlations served as reliability scores to threshold the data, ensuring consistent analysis across trials and channels.

\paragraph{Training parameters}
\label{sec:hyperparams}
The dataset was divided into 22,348 training samples and 100 test samples. We utilized the ADAM optimizer and set the learning rate to 0.0078, with beta coefficients at 0.5 and 0.999 to optimize convergence. The loss function weights were configured as follows: discriminator loss (\(\alpha_{\text{discr}}\)) at 0.01, VGG feature loss (\(\beta_{\text{vgg}}\)) at 0.69, and L1 pixel-wise loss (\(\beta_{\text{pix}}\)) at 0.3, balancing the model's sensitivity to different types of error. Training was conducted over 100 epochs on a Quadro RTX 6000 GPU, utilizing approximately 10000 MiB of GPU memory. 

\paragraph{Baseline Model}
\label{sec:baseline}
The original model described by Shen et al.~\citep{Shen2019end} incorporates a generator composed of three fully connected layers and six upconvolution layers aimed at producing reconstructed images. The model also integrates a VGG-based comparator (commonly referred to as VGG loss) and a discriminator. Upon implementing this model using the same dataset as our study, the outputs were unsatisfactory, predominantly yielding only yellow images. To address this, we enhanced the generator's complexity, diverging from the Shen et al. setup while retaining certain foundational aspects. Unlike the original model that used an AlexNet comparator, we used a VGG feature loss for extracting features from the final convolutional layer of VGG11 in order to perform qualititative analysis with AlexNet.

\paragraph{Model performance evaluation}
The performance of the models were quantified by correlating features of the reconstructions with the stimuli. Specifically, features from reconstructed images and original stimuli are extracted using a pretrained AlexNet model at conv1, conv2, conv3, conv4, conv5 FC6, FC7, and FC8. Subsequently, each feature layer is analyzed using the Pearson correlation method to assess the correlation between the features of the reconstructions and those of the original stimuli.

\section{Results}

\subsection{Multiple features are needed for better reconstruction}

\begin{figure}[!ht]
  \centering
  \includegraphics[width=0.8\textwidth]{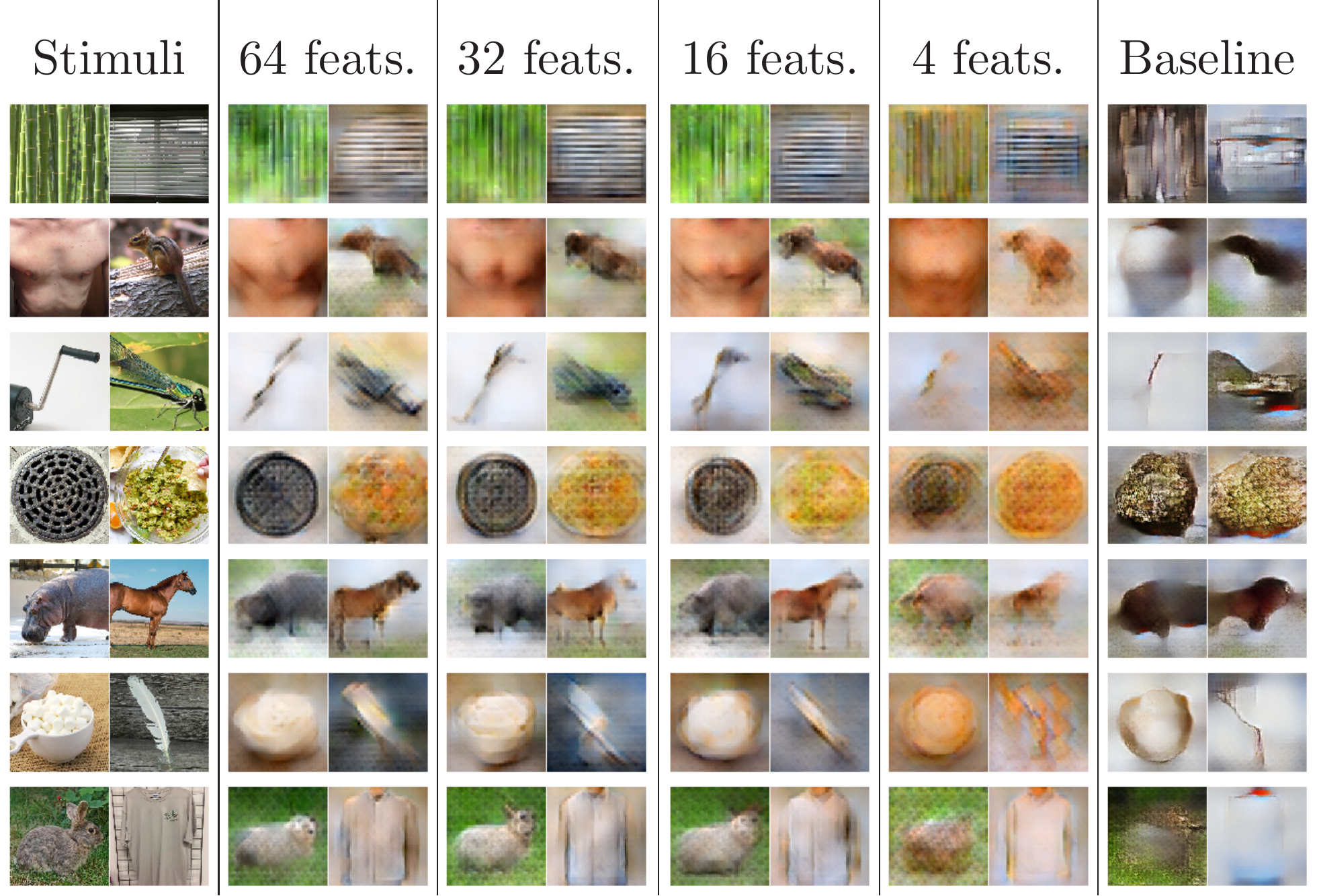}
  \caption{Reconstructions of end-to-end training with various number of feature channels. Baseline is based on the end-to-end model of Shen et al.~\citep{Shen2019end}.}
  \label{fig:features}
\end{figure}

In our investigation, we explored the impact of varying the number of dedicated feature channels within our model, specifically training with 4, 16, 32, and 64 features, to determine if increasing feature separability leads to enhanced model performance. Given the convolutional nature of the model, there is an inherent preservation of spatial integrity, as the convolution operation utilizes relevant spatial information for computation. Our hypothesis posited that while an increase in feature channels could introduce redundancy and augment the complexity of the model due to an increase in learnable parameters, it would nevertheless result in improved reconstruction capabilities. Supporting our hypothesis, the results clearly demonstrate that a greater number of features significantly enhances reconstruction quality as shown in Figure \ref{fig:features}.

% \textpm
\begin{table}[ht]
  \caption{Comparison of reconstruction quality across layers and feature counts.}
  \label{table:feature-comparison}
  \centering
  \begin{tabular}{lccccc}
    \toprule
     & \multicolumn{4}{c}{Number of learned features} \\
    \cmidrule(r){2-5}
     Layer      & 4      & 16     & 32     & 64  & Baseline    \\
    \midrule
    conv1  & 0.238 & 0.289 & 0.300 & \textbf{0.302} & 0.267 \\
    conv2  & 0.172 & 0.219 & 0.227 & \textbf{0.231} & 0.221 \\
    conv3  & 0.273 & 0.343 & 0.346 & \textbf{0.355} & 0.326 \\
    conv4  & 0.289 & 0.330 & 0.345 & \textbf{0.353} & 0.316 \\
    conv5  & 0.207 & 0.262 & \textbf{0.276} & 0.275 & 0.203 \\
    FC6    & 0.213 & 0.294 & 0.308 & \textbf{0.310} & 0.235 \\
    FC7    & 0.394 & 0.488 & 0.504 & \textbf{0.505} & 0.434 \\
    FC8    & 0.420 & 0.517 & \textbf{0.531} & 0.515 & 0.446 \\
    \bottomrule
  \end{tabular}
\end{table}

\begin{figure}[!ht]
  \centering
  \includegraphics[width=\textwidth]{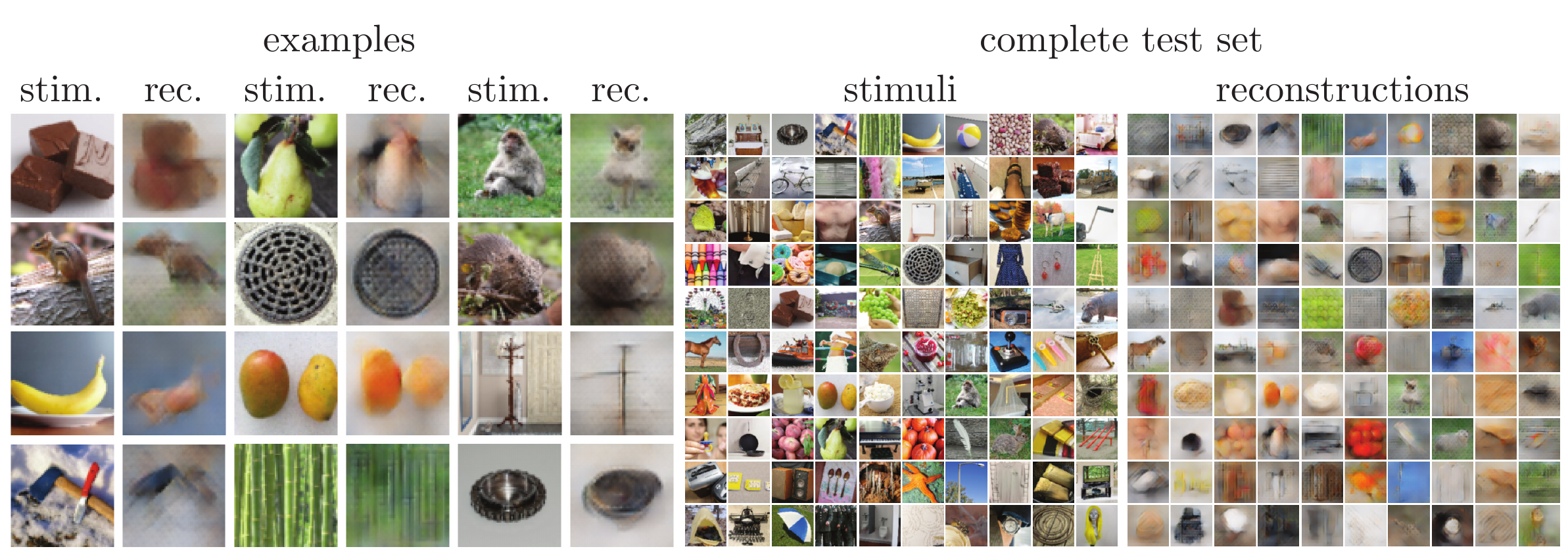}
\caption{Reconstructions from our IRFA end-to-end reconstruction model. Example reconstructions from the test-set (left) and all the reconstructions from the test-set (right).}
  \label{fig:all_recons}
\end{figure}

Table \ref{table:feature-comparison} supports this observation quantitatively and underscores the importance of incorporating feature inverse receptive fields (IRFs) alongside spatial IRFs. The Pearson correlation test yielded p-values \( \ll 0.01 \), indicating a statistically significant relationship. Consequently, subsequent experiments detailed in this paper will refer to the model configured with 64 feature channels. Figure~\ref{fig:all_recons} depicts reconstructions obtained using 64 feature channels. Stimuli with high-frequency features tend to have more reconstruction errors, likely due to the distribution of recording sites across visual areas. The model relies heavily on V1, which has significantly more recording sites than V4 and IT. Since V1 processes fine-grained spatial details, while V4 and IT integrate more complex features, the limited coverage in these higher visual areas may reduce the model's ability to capture high-frequency details accurately across the entire visual hierarchy. This suggests that reconstruction accuracy may vary with stimulus frequency due to limitations in recording site coverage.

\begin{figure}[!ht]
  \centering
  \includegraphics[width=\textwidth]{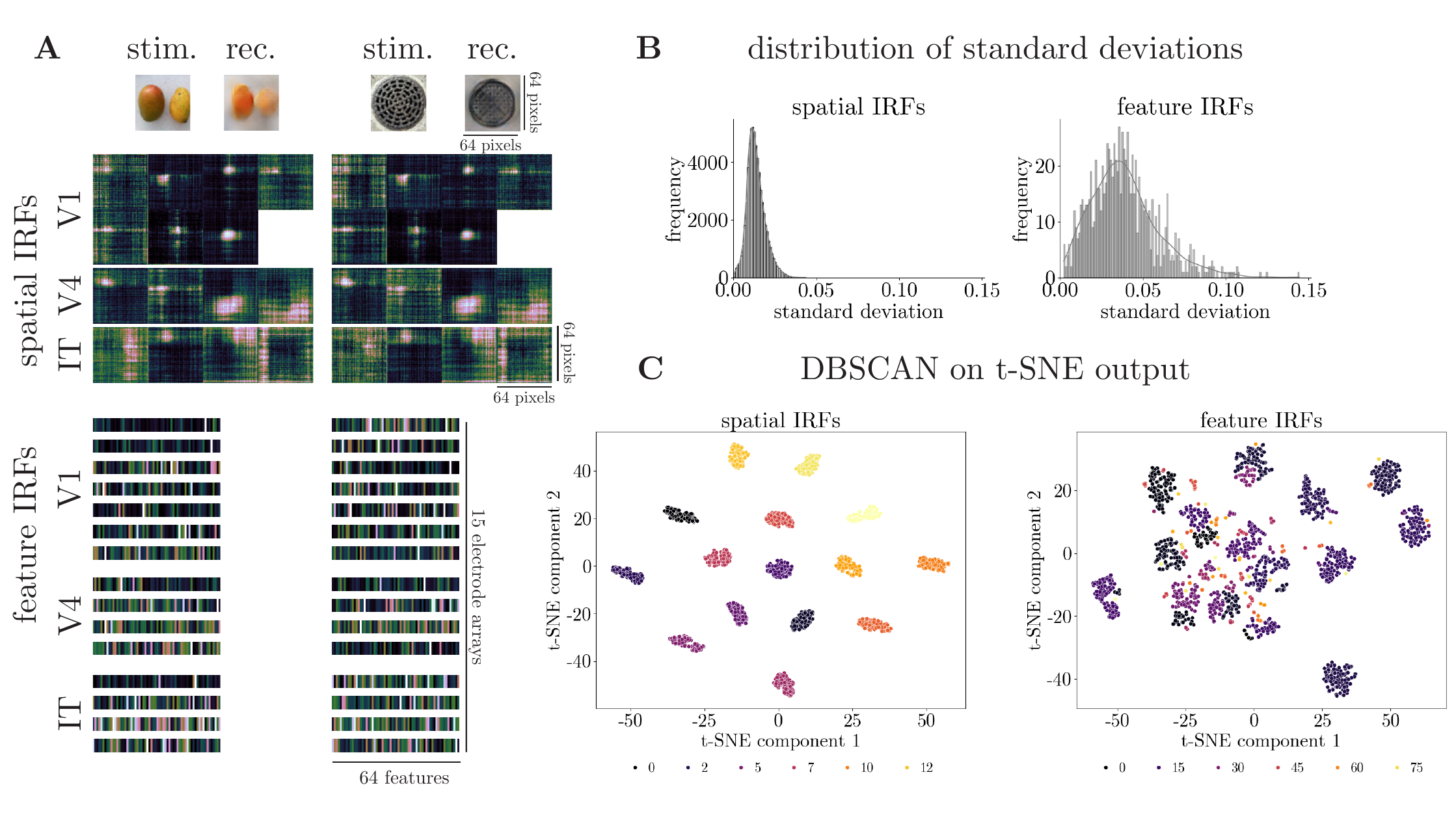}
\caption{\textbf{A}. Depiction of learned spatial inverse receptive fields (spatial IRFs) and feature inverse receptive fields (feature IRFs) assigned to the reconstructed test images (stim. / rec.) by our trained reconstruction model. \textbf{B}. This panel shows the frequency distribution of specific IRFs across test stimuli, where frequency represents the number of times each IRF appears across different stimuli, illustrating the consistency of spatial IRFs across stimuli compared to feature IRFs. The labels correspond to distinct brain regions associated with 15 electrodes. \textbf{C}. Application of density-based spatial clustering of applications with noise (DBSCAN) to t-distributed stochastic neighbor embedding (t-SNE) data identifies 576 clusters in spatial IRFs, corresponding to the number of electrode channels, and 1113 clusters in feature IRFs, representing groups of features across electrode channels responding to 100 stimuli.}

  \label{fig:IRFs}
\end{figure}

\subsection{Consistency and variability in receptive fields}
Despite the spatial inverse receptive fields (IRFs) not being explicitly trained on real data, they emerged naturally during the image reconstruction process and displayed remarkable consistency across varying stimuli within each of the 15 electrode channels, as illustrated in Figures \ref{fig:IRFs}A and \ref{fig:IRFs}B. Notably, spatial IRFs associated with V1 neurons exhibited greater locality compared to those from V4 and IT regions, consistent with established neuroanatomical insights. This suggests that the model-generated spatial IRFs can accurately represent the spatial receptive fields of the electrode arrays.

Further analysis involved assessing the variability of both spatial and feature-based receptive fields (SRFs and FRFs) by calculating their standard deviations across all test stimuli. Represented in arrays for the 15 electrodes with respective dimensions, these fields were analyzed to determine their stability or variability, which was visualized through histograms of these standard deviation values.

Distinctively, feature IRFs refer to the variability of features, or grid channels, that selectively encode specific components of the visual stimulus at each spatial location. This variability across stimuli indicates how feature-based selective attention shifts dynamically to capture different aspects of the input, shown in Figure \ref{fig:IRFs}A (feature IRFs). Unlike the spatial receptive fields, feature IRFs showed significant variability across stimuli, with higher standard deviations indicating dynamic shifts in attention to different features (see Figure \ref{fig:IRFs}B). This variability suggests that feature-based selective attention plays a role in our reconstruction process, as the model adapts dynamically to different stimuli by adjusting feature attention. These findings underscore the novelty and significance of our model in revealing intricate patterns of neural encoding and attention dynamics.

\subsection{Data-driven clustering shows more clusters for features than for spatial IRFs}

In our study, we applied t-distributed stochastic neighbor embedding (t-SNE)~\cite{Maaten2008} with a perplexity setting of 30 for dimensionality reduction, followed by density-based spatial clustering of applications with noise (DBSCAN)~\cite{Ester1996} with an epsilon value of 1.5, to visualize the clustering patterns of spatial and feature-specific attention mechanisms encoded by neural activity. The initial multidimensional array of learned inverse receptive fields (IRFs) (\(100 \times 15 \times 64 \times 64 \times 64\)) was preprocessed by averaging across the feature dimension to produce spatial IRFs (\(100 \times 15 \times 64 \times 64\)), and across spatial dimensions to derive feature IRFs (\(100 \times 15 \times 64\)). These adjustments ensured each set reflected the attentional weights neural recordings assigned across 15 electrode arrays.

After flattening these arrays, t-SNE effectively reduced the high-dimensional data to a two-dimensional space, facilitating a subsequent DBSCAN analysis to determine clustering patterns. As shown in Figure~\ref{fig:IRFs}C, this analysis revealed that while spatial IRFs formed consistent clusters corresponding to electrode arrays, feature IRFs exhibited a significantly higher number of clusters. This disparity highlights the neurons' varied capacity to differentiate and prioritize among diverse features within visual stimuli. The results from this data-driven clustering provide critical insights into how neurons integrate and prioritize information, reflecting the complex interplay of spatial and feature-specific attention mechanisms in visual processing.

\section{Discussion}

In this study, we explored how neurons contribute to perception by leveraging receptive fields in visual neuroscience. Traditionally, receptive fields describe how specific neurons are tuned to regions of the visual field. Extending this concept, our model introduces inverse receptive fields, where each pixel in a visual stimulus is influenced by a subset of neurons. Incorporating a trainable attention component to learn feature-based attention significantly enhanced the reconstruction of naturalistic images compared to models relying solely on spatial attention mechanisms.

Our results show consistent spatial selectivity in inverse receptive fields (IRFs), modeled without fixation maps or retinotopy data. This suggests that these maps encapsulate crucial information about the locations neurons focus on during visual processing. While spatial receptive fields were consistent, variability in feature selectivity aligned with the biased competition theory proposed by Desimone and Duncan \cite{desimone1995neural}, supporting feature selection as a fundamental aspect of neuronal activity.

In literature, feature-based attention modulates neuronal activity across visual areas such as V4, MT, and IT \cite{maunsell2006feature}. Traditionally, spatial and feature-based attention are seen as facets of a single neurobiological mechanism, with spatial location considered a feature itself. Despite the visuotopic organization of most visual structures, feature attention impacts neuronal responses across the entire visual field. Our model's ability to distinguish between spatial and feature receptive fields provides a novel tool for dissecting these neurophysiological mechanisms.

While recent image reconstruction models, such as diffusion models, produce high-resolution outputs, our study prioritizes spatial accuracy and interpretability in an end-to-end framework. Diffusion models, though effective in generating detailed images, have objectives distinct from ours, given their reliance on separate generative processes rather than latent representation learning. We compared our model to an end-to-end baseline derived from Shen et al. (2019) \cite{Shen2019end}, aligned with our goals of spatially accurate reconstructions.

Compared to other end-to-end reconstruction approaches, our model builds on the Brain2Pix framework \cite{le2022brain2pix}, a state-of-the-art model in neural decoding, and extends beyond early methods like Miyawaki et al. (2008) \cite{miyawaki2008visual}. While Miyawaki et al. (2008) was foundational, it lacks the end-to-end learning capabilities and attention mechanisms of recent models. Shen et al. (2019) \cite{Shen2019end}, which we use as a benchmark, provides an updated model with deeper architectures and more robust feature representations. By integrating feature-based attention and inverse receptive fields, our model improves on these approaches, offering a more accurate and interpretable reconstruction of neural data.

\subsection{Broader impact} \label{sec:broaderimpacts}
Neural decoding models, particularly those tasked with reconstructing naturalistic images, significantly deepen our understanding of the relationship between neural activity and the stimuli that evoke it. These models hold substantial promise for applications such as visual neuroprosthetics, aiming to restore lost visual experiences. However, their deployment must be approached with caution due to the inherent complexities of brain function and environmental interactions. Human engagement with the environment extends beyond simple perception and attention, involving complex behaviours and neuroplastic adaptations that generalized models, based solely on image data, may fail to capture. For example, visual feedback from motor outputs presents a layer of complexity not accounted for in static image training. Moreover, using these models to map stimulation locations for neuroprosthetic purposes may not yield accurate replications of natural neural responses, as the act of stimulating brain areas does not mimic the dynamics of natural sensory recording. Such discrepancies highlight the critical need for nuanced model development and the cautious interpretation of model outputs in practical applications.

Further insights into image reconstruction from brain activity and its influence on perception could also advance learning processes by improving how we capture and maintain attention. Nonetheless, there is a potential risk that these advancements could be misused to target individuals without their consent, raising ethical concerns. Although the feasibility of exploiting this technology is currently limited—primarily due to the rarity of single-neuron recordings in humans—the ethical implications warrant careful consideration as the field advances. 

\subsection{Limitations} 

The applicability of our results to non-invasive techniques, such as fMRI, remains unclear due to their lower signal-to-noise ratio and limited spatial resolution compared to electrode arrays. Additionally, replicability may be impacted by uneven electrode coverage, as our setup has dense coverage in V1 but fewer sites in V4 and IT, potentially affecting reconstructions of high-frequency features. While our model minimizes these effects with spatial and feature-based attention, further work is needed to confirm robustness across different configurations and recording conditions.

\medskip

{
\small

\bibliographystyle{unsrt}

}

\newpage
\appendix

\section{Appendix / Supplemental material}
\subsection{Model ablation}
\label{appendix:ablation}
\begin{figure}[!ht]
  \centering
  \includegraphics[width=\linewidth]{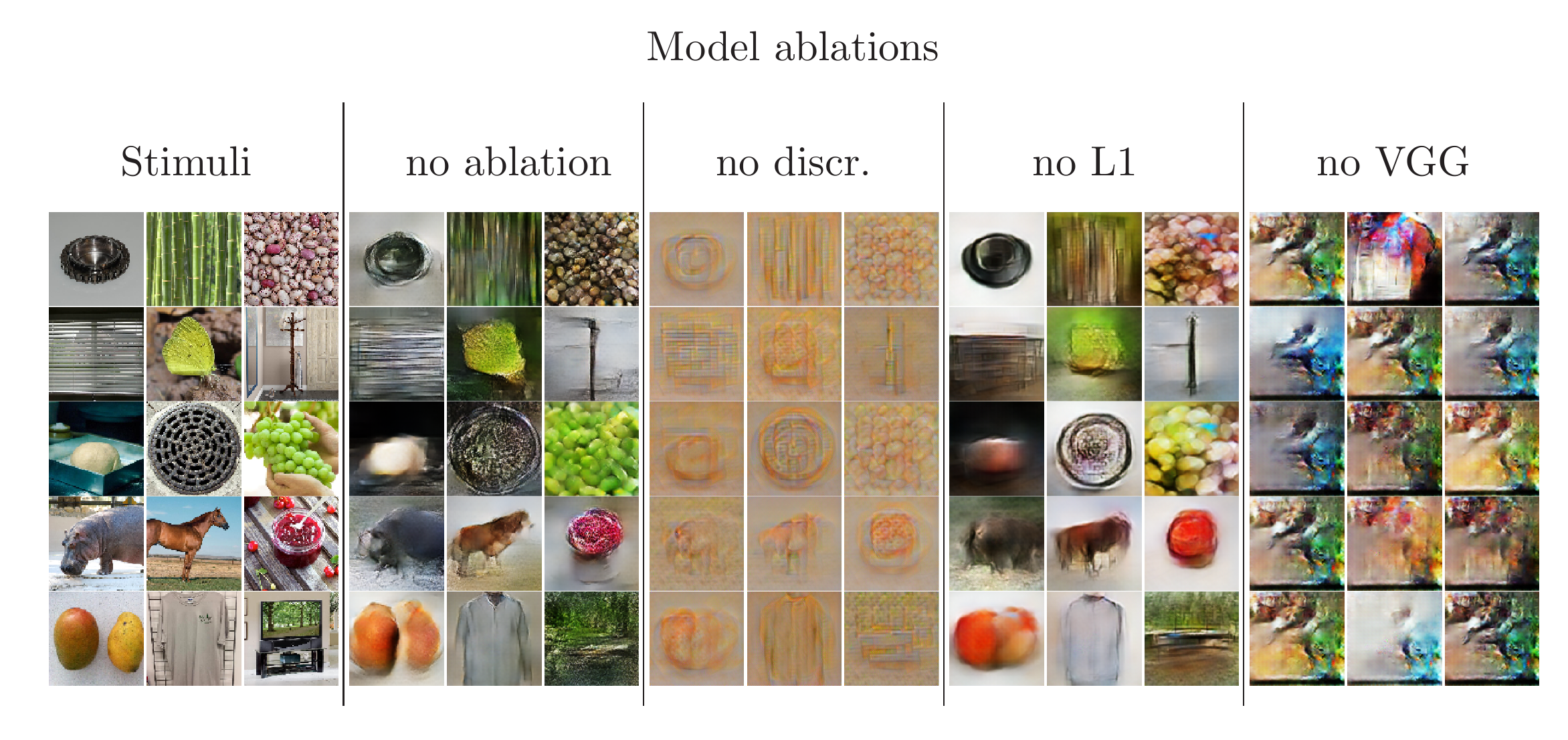}
  \caption{Ablation study results illustrating the effect of removing each core component of the framework (discriminator, VGG, and L1 losses) on reconstruction quality, excluding the attention component. The figure shows reconstruction outputs under four conditions: (1) full model with all components, (2) model without the discriminator, (3) model without the VGG loss, and (4) model without the L1 loss. Removing any component reduces reconstruction fidelity, underscoring the importance of each component to the model’s performance.}
  \label{fig:ablation_results}
\end{figure}

To assess the contributions of the core components of our framework, we conducted an ablation study by selectively removing the discriminator, VGG, and L1 losses (with the attention component excluded for these tests). The discriminator enhances realism by reinforcing structural fidelity in the reconstructions, the VGG loss contributes to perceptual quality by preserving high-level feature consistency, and the L1 loss improves pixel-wise accuracy. The ablation results, as shown in \ref{fig:ablation_results} demonstrate that each component plays a critical role in achieving high-quality reconstructions. Removing any of these losses led to degraded outputs, validating their necessity within the framework. These findings align with prior studies that establish these components as integral to robust neural decoding and image reconstruction.

\end{document}